\documentclass[conference]{IEEEtran}
\IEEEoverridecommandlockouts
\usepackage{cite}
\usepackage{amsmath,amssymb,amsfonts}
\usepackage{graphicx}
\usepackage{textcomp}
\usepackage{xcolor}
\def\BibTeX{{\rm B\kern-.05em{\sc i\kern-.025em b}\kern-.08em
    T\kern-.1667em\lower.7ex\hbox{E}\kern-.125emX}}

\usepackage{algorithm}
\usepackage{algpseudocode}
\usepackage{amsfonts}
\usepackage{subcaption}

\usepackage{cleveref}

\usepackage{caption}
\usepackage{enumitem}

\usepackage{url}

\usepackage{titlesec}
\titlespacing\section{0pt}{0.5\baselineskip}{0.25\baselineskip}
\titlespacing\subsection{0pt}{0.25\baselineskip}{0.15\baselineskip}
\titlespacing\subsubsection{0pt}{0.15\baselineskip}{0.1\baselineskip}

\begin{document}

\title{AdvQuNN: A Methodology for Analyzing the Adversarial Robustness of Quanvolutional Neural Networks\\
\vspace{-10pt}
}


\author{\IEEEauthorblockN{Walid El Maouaki\textsuperscript{1}, 
Alberto Marchisio\textsuperscript{2,3}, 
Taoufik Said\textsuperscript{1},
Mohamed Bennai\textsuperscript{1},
and Muhammad Shafique\textsuperscript{2,3}
}
\IEEEauthorblockA{
  \textsuperscript{1}Quantum Physics and Magnetism Team, LPMC, Faculty of Sciences Ben M’Sik, Hassan II University of Casablanca, Morocco\\
  \textsuperscript{2}eBrain Lab, Division of Engineering, New York University Abu Dhabi (NYUAD), Abu Dhabi, UAE\\
  \textsuperscript{3}Center for Quantum and Topological Systems (CQTS), NYUAD Research Institute, NYUAD, Abu Dhabi, UAE\\
  email: walid.elmaouaki-etu@etu.univh2c.ma, alberto.marchisio@nyu.edu, taoufik.said@univh2c.ma,\\
  mohamed.bennai@univh2c.ma, muhammad.shafique@nyu.edu
}
\vspace{-30pt}
}

\maketitle

\begin{abstract}
Recent advancements in quantum computing have led to the development of hybrid quantum neural networks (HQNNs) that employ a mixed set of quantum layers and classical layers, such as Quanvolutional Neural Networks (QuNNs). While several works have shown security threats of classical neural networks, such as adversarial attacks, their impact on QuNNs is still relatively unexplored. This work tackles this problem by designing AdvQuNN, a specialized methodology to investigate the robustness of HQNNs like QuNNs against adversarial attacks. It employs different types of Ansatzes as parametrized quantum circuits and different types of adversarial attacks. This study aims to rigorously assess the influence of quantum circuit architecture on the resilience of QuNN models, which opens up new pathways for enhancing the robustness of QuNNs and advancing the field of quantum cybersecurity. Our results show that, compared to classical convolutional networks, QuNNs achieve up to 60\% higher robustness for the MNIST and 40\% for FMNIST datasets. 

\end{abstract}



\begin{IEEEkeywords}
Quantum machine learning, Quanvolutional Neural Networks, Quantum Computing, Deep Neural Networks, Convolutional Neural Networks, Adversarial Attacks, Adversarial Robustness
\end{IEEEkeywords}



\maketitle
\pagestyle{plain}

\section{Introduction}

Driven by the emergent trends of quantum computing and the massive learning capabilities of Deep Neural Networks (DNNs), Hybrid Quantum Neural Networks (HQNNs) have emerged as high-accurate models to conduct machine learning tasks on currently available quantum devices~\cite{zaman2023survey, markidis2023programming}. Due to the reliability constraints of the current existing quantum circuits that limit their sizes, HQNNs combine resource-constrained quantum circuits in quantum layers with classical layers implemented on quantum and classical computers. A common HQNN model is the Quanvolutional Neural Networks (QuNNs)~\cite{henderson2020quanvolutional} that implement a quanvolutional layer followed by classical layers. 

\subsection{Limitations of State-Of-The-Art and Main Contributions}

Before employing QuNNs in safety-critical systems, it is important to analyze their robustness to different vulnerability threats. While the security of classical DNNs has been extensively studied~\cite{guesmi2023physical}, the security of QuNNs has not been analyzed systematically. The recent work in~\cite{sooksatra2021evaluating} has compared QuNNs with analogous classical networks and the work in~\cite{zaman2024studying} studied the impact of quantum hyperparameters on the accuracy of QuNNs. However, they did not account for a range of perturbation strength values, while \textit{we believe that a more thorough comparison necessitates evaluating the robustness of the models across a spectrum of increased perturbation strengths}. Moreover, the work in~\cite{huang2023image} analyzed the effect of different sizes of the kernel and different numbers of filters in QuNNs, which is not sufficient as it is a classical-inspired permutation, and does not fully encapsulate the quantum-specific dynamics, while \textit{in our work we investigate the impact of the type of quantum circuit (Ansatz), offering a more inherent quantum perspective.}

The works in~\cite{akter2023exploring} and~\cite{wendlinger2024comparative} identify significant adversarial vulnerabilities in conventional neural networks and quantum neural networks, noting that while quantum models exhibit higher resilience and intrinsic robustness due to their architectural properties, they still face considerable adversarial threats. However, neither study delves into the specific design and optimization of quantum circuit architectures to evaluate and enhance robustness. \textit{Our work directly addresses this gap by presenting an empirical analysis of various Ansatz architectures in QuNNs}.
Additionally, the works in~\cite{du2021quantum} and~\cite{huang2023certified} demonstrate that introducing noise, such as depolarization and random rotation noise, enhances the robustness of quantum classifiers against adversarial attacks while satisfying quantum differential privacy. In contrast, \textit{our work studied the impact of entanglement in the quantum circuit architecture of QuNNs where we demonstrated that a more interconnected circuit can enhance the model resilience}.
The work in~\cite{liu2020vulnerability} provides an overview of the vulnerabilities and potential robustness of QML against adversarial attacks, emphasizing the intrinsic properties of Hilbert space dimensions and their implications for adversarial robustness. Building upon this work, \textit{our study explores the impact of Hilbert space expressibility and entanglement strength properties from the circuit architecture perspective on the QuNN resilience}.





Selecting a robust architecture for the quanvolutional filter circuit, or the appropriate set of quantum gates, is not a straightforward task. It often requires empirical investigation as the ideal circuit configuration is not readily apparent without practical trials. Moreover, searching for the best possible circuit design can sometimes demand more computational resources than using existing circuit designs as a starting point or inspiration. The suitability of a given architecture is also influenced by the specific characteristics of the dataset in question. In this work, we explore and assess several Ansatz architectures, identifying, testing, and comparing with efficient parametric circuit configurations tailored to our particular datasets, thereby enhancing the robustness of QuNN models. Through this evaluation process, we aim to establish a balance between the design of new quantum circuits in the quanvolutional layer and the robustness of the QuNN models.

\vspace{3pt}

\noindent
\textbf{Our Novel Contributions}

\vspace{3pt}

\begin{figure}[!ht]
  \centering
  \includegraphics[width=0.9\linewidth]{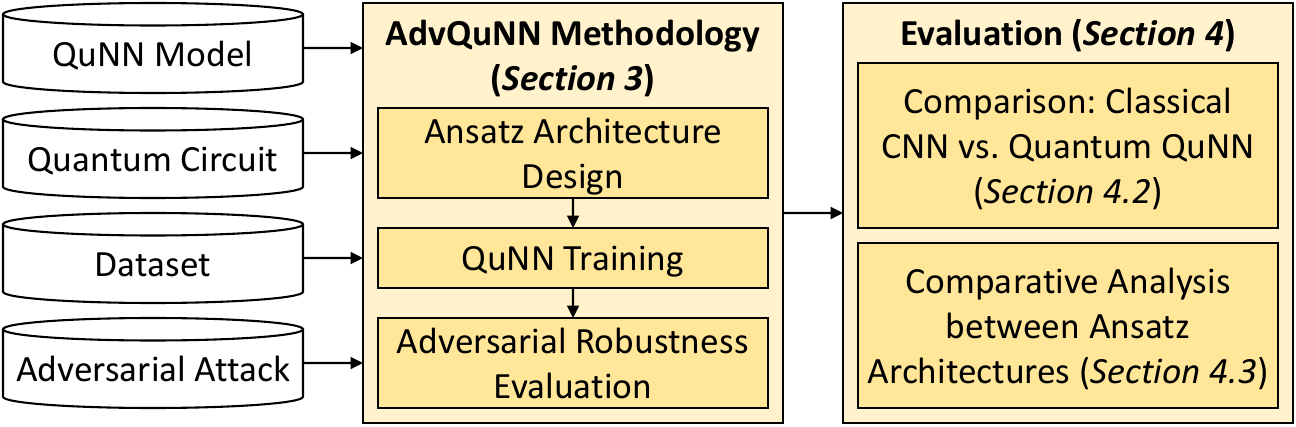}
  \caption{Overview of our novel contributions in this work.}
  \label{fig:novel_contributions}
  \vspace{-5pt}
\end{figure}

\noindent
In this paper, we present \textit{AdvQuNN}, a methodology for analyzing the adversarial robustness of QuNNs (see \Cref{fig:novel_contributions}). Our contributions to this paper are:

\begin{itemize}[leftmargin=*]
    \item We design the AdvQuNN methodology to test different types of adversarial attacks on QuNNs and classical DNN models, varying the quantum circuit type. (\textbf{\Cref{sec:methodology}})
    \item We compare the robustness of QuNNs to classical CNNs against various types of adversarial attacks. (\textbf{\Cref{subsec:comparison_Classical_Quantum}})
    \item We evaluate the impact of the Ansatz architecture on the QuNN robustness (\textbf{\Cref{Comparative_Analysis_between Ansatz_Architectures}})
\end{itemize}

\section{Background and Related Work}



\subsection{Classical Convolutional Neural Networks}
Convolutional Neural Networks (CNNs)~\cite{o2015introduction, alzubaidi2021review}, particularly effective in visual tasks~\cite{li2021survey} like image classification~\cite{rawat2017deep}, object detection~\cite{zhiqiang2017review}, and semantic segmentation~\cite{guo2018review}, rely on convolutional layers to extract spatial features. These layers apply filters to input images, producing feature maps that represent detected features. The convolution operation is mathematically defined as $F(i,j)=(X * K)(i, j)=\sum_m \sum_n X(m,n) K(i-m, j-n)$, where $F$ is the feature map, $X$ is the input image, and $K$ is the kernel. Subsequent layers combine these features into more complex patterns, following $F^l=\sigma\left(\sum_{k=1}^K F_k^{l-1} * K_k^l+b_k^l\right)$, with $\sigma$ as the activation function.

\subsection{Quanvolutional Neural Networks}
Quanvolutional Neural Networks (QuNNs)~\cite{henderson2020quanvolutional} represent an intersection between the domains of quantum computing and deep learning that provides an advanced approach to process and understand data. QuNN models substitute the conventional convolutional filter in the CNN with a quantum filter known as the quanvolutional filter. Like classical convolutional layers, the quanvolutional layer applies an operation analogous to an $m \times m$ filter, but it does so by executing an $m^2$-qubit quantum circuit (if we use direct pixel encoding as in this work) to produce a downsampled output image. 
The quanvolutional layer includes three components: data encoding, a parameterized unitary $U(\theta)$, and measurement $M$.

\textbf{Data encoding.} The input data to the quantum layer, consisting of $n$ features, is encoded into an $n$-qubit Hilbert space using quantum data encoding techniques~\cite{larose2020robust}. In this work, we used the angle encoding method where each classical data feature maps to a rotation angle, thus preserving data structure in the quantum circuit. The quantum state of an angle encoded input data $x=[x_0, x_1, \dots, x_n]^{\intercal} \in \mathbb{R}^n$ is represented as

\vspace{-5pt}
\begin{equation}
|\psi_x\rangle=\bigotimes_i^n R\left(\mathbf{\phi}_i\right)\left|0^n\right\rangle
\vspace{-5pt}
\end{equation}

where $R$ can be one of $R_x$, $R_y$, $R_z$. \textit{In this work, we chose $R_y$ due to its ability to affect both the amplitude and phase of the qubit state for a more expressive data encoding in the quantum state space.} The original pixel values $x \in [0,1]$ is scaled to $\phi \in [0, \pi]$, hence $\mathbf{\phi}_i=\pi \times x_i$.

\textbf{Parameterized unitary $U(\theta)$.} A unitary operator $U$, represented by a $2^n \times 2^n$ matrix, configures an $n$-qubit quantum circuit using single-qubit gates (e.g., ${R_x, R_y, R_z, H, \dots}$) and two-qubit gates (e.g., ${CNOT, CR_x, CR_y, CR_z, \dots}$). These gates enable the creation of various unitary operations $U_i$, each forming a distinct quanvolutional filter with specific gate sequences and entanglement patterns. Parameters for these filters can be iteratively updated with classical layer parameters or initialized randomly. \textit{In this study, our methodology focuses on the impact of optimizing circuit configuration rather than its parameters, hence, we opt for random initialization. This also will allow us to reduce computational demands and simplify the quantum circuit configuration process.} The selected parametric circuit $U(\theta)$ is applied to the initial state as follows:


\begin{equation}
    |\psi_x (\theta)\rangle=U(\theta)|\psi_x\rangle
\end{equation}

\textbf{Measurement $M$.} Measuring the quantum state $\left|\psi_x(\theta)\right\rangle$ with observable $M=Z^{\otimes n}$ ($Z$ is the Pauli-$Z$ matrix) yields probabilistic outcomes, where the probability $p_i$ represents the likelihood of observing the eigenvalues $\lambda_i$ of $M$. This facilitates the calculation of expectation values for $M$ with respect to the quantum state $\left|\psi_x(\theta)\right\rangle$ as:
\begin{equation}
    \langle M \rangle = \langle \psi_x(\theta)| M |\psi_x(\theta) \rangle= \frac{1}{S} \sum_i p_i \lambda_i
\end{equation}

The expectation value, derived from $S$ measurements, forms a feature map for subsequent processing by classical layers.

\begin{figure*}[h]
    \centering
    \includegraphics[width=0.8\linewidth]{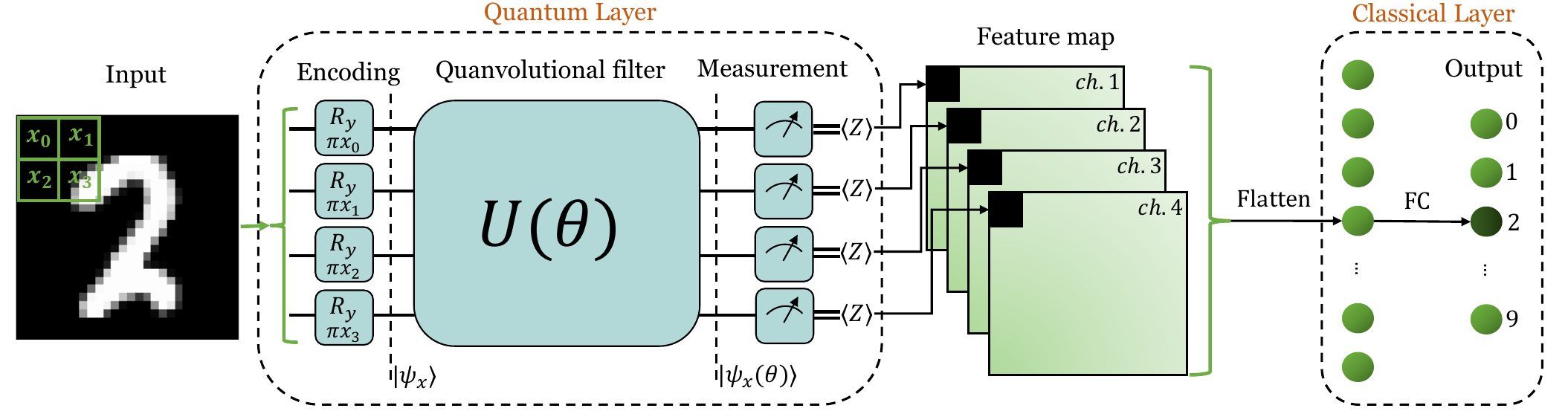}
    \caption{Quanvolutional Neural Network architecture. A 2 by 2 block of the input image of digit $2$ is encoded through a of rotation gates $R_y$, producing the state $|\psi_x\rangle$. This state is quanvolved with a parameterized unitary gate  $U(\theta)$. Subsequently, we measure the state $|\psi_x(\theta)\rangle$ and compute the expectation value across all qubits, and assign values to four channels, forming a feature map. Finally, a sequence of fully connected classical layers then processes this map for classification.}
    \label{fig:QNN_architecture}
\end{figure*}

Figure~\ref{fig:QNN_architecture} shows an example of a QuNN architecture applied to the MNIST dataset. A $2\times 2$ square of our input image is first encoded with $R_y$ gates into the $2\times 2$ quanvolutional kernel of $4$ qubits. This data then undergoes transformation by a unitary $U(\theta)$. Finally, we measure the expectation value of the Pauli-$Z$ observable for each qubit. Each expectation value contributes to a separate channel in the resulting pixel. Therefore, with four qubits, we obtain four channels. These channels are subsequently flattened and input into a classical dense layer, which determines the likelihood of each digit.

\subsection{Adversarial Attacks}
In Quantum Machine Learning (QML), particularly with quanvolutional neural networks,  the issue of adversarial attacks presents a dual challenge of vulnerability and potential robustness. The works in~\cite{liu2020vulnerability} and~\cite{lu2020quantum} reveal that these networks, whether they process classical or quantum input data, share vulnerabilities with classical networks, while also presenting a trade-off between algorithmic security and quantum advantages. Conversely, studies in~\cite{guan2021robustness} and~\cite{west2023benchmarking} highlight an intrinsic robustness of quantum models to such attacks. The works in~\cite{sooksatra2021evaluating},~\cite{suryotrisongko2022adversarial},~\cite{ren2022experimental}, and~\cite{dowling2024adversarial} further confirm the superior performance of QuNNs over classical networks in terms of accuracy and resilience against adversarial attacks both empirically and theoretically. \textit{Our study focuses on investigating a novel approach to evaluate and enhance the robustness of QuNNs.} 
The attack algorithms used in this work, FGSM, PGD, and MIM, are shown in Algorithms~\ref{alg:FGSM},~\ref{alg:PGD}, and~\ref{alg:MIM}, respectively.

\begin{figure}[h]
\begin{algorithm}[H]
\caption{FGSM Attack}
\label{alg:FGSM}
\begin{algorithmic}[1]
 \renewcommand{\algorithmicrequire}{\textbf{Input:}}
 \renewcommand{\algorithmicensure}{\textbf{Output:}}
\Require Model $model$, Data $X$ of size $M$, True Labels $y$, Perturbation size $\epsilon$
\Ensure Adversarial examples $\hat{X}$

\Function{FGSM\_Attack}{$model, X, y, \epsilon$}
    \State $gradients \gets$ \Call{Compute\_Gradient}{$model, X, y$}
    \State $sign\_gradients \gets \text{sign}(gradients)$
    \State $\hat{X} \gets X + \epsilon \cdot sign\_gradients$
    \State $\hat{X} \gets \text{clip}(\hat{X}, 0, 1)$
    \State \Return $\hat{X}$ of size $M$
\EndFunction

\Function{Compute\_Gradient}{$model, X, y$}
    \State \textbf{with} GradientTape() \textbf{as} tape:
    \State \hspace{1em} tape.watch($X$)
    \State \hspace{1em} prediction $\gets$ model($X$)
    \State \hspace{1em} loss $\gets$ sparse\_categorical\_crossentropy($y$, prediction)
    \State gradients $\gets$ tape.gradient(loss, $X$)
    \State \Return $gradients$
\EndFunction

\end{algorithmic}
\end{algorithm}
\vspace{-5pt}
\end{figure}
\begin{figure}[h]
\begin{algorithm}[H]
\caption{PGD Attack}
\label{alg:PGD}
\begin{algorithmic}[1]
 \renewcommand{\algorithmicrequire}{\textbf{Input:}}
 \renewcommand{\algorithmicensure}{\textbf{Output:}}
\Require Model $model$, Data $X$ of size $M$, True Labels $y$, Perturbation size $\epsilon$, Iterations $k$, Step size $\alpha$
\Ensure Adversarial examples $\hat{X}$

\Function{PGD\_Attack}{$model, X, y, \epsilon, k, \alpha$}
    \State $\hat{X} \gets X$
    \For{$i = 1$ to $k$}
        \State \Call{FGSM\_Attack}{$model, \hat{X}, y, \epsilon$}
        \State $\hat{X} \gets \text{clip}(\hat{X}, X - \epsilon, X + \epsilon)$
        \State $\hat{X} \gets \text{clip}(\hat{X}, 0, 1)$
    \EndFor
    \State \Return $\hat{X}$ of size $M$
\EndFunction

\end{algorithmic}
\end{algorithm}
\vspace{-5pt}
\end{figure}

\begin{figure}[h]
\begin{algorithm}[H]
\caption{MIM Attack}
\label{alg:MIM}
\begin{algorithmic}[1]
 \renewcommand{\algorithmicrequire}{\textbf{Input:}}
 \renewcommand{\algorithmicensure}{\textbf{Output:}}
\Require Model $model$, Data $X$ of size $M$, True Labels $y$, Perturbation size $\epsilon$, Iterations $k$, Step size $\alpha$, Decay factor $\mu$
\Ensure Adversarial examples $\hat{X}$

\Function{MIM\_Attack}{$model, X, y, \epsilon, k, \alpha, \mu$}
    \State $\hat{X} \gets X$
    \State $g \gets 0$ \Comment{Initialize the momentum term}
    \For{$i = 1$ to $k$}
        \State $gradients \gets$ \Call{Compute\_Gradient}{$model, \hat{X}, y$}
        \State $g \gets \mu \cdot g + \frac{gradients}{\|gradients\|_1}$ \Comment{Update momentum term}
        \State $sign\_gradients \gets \text{sign}(g)$
        \State $\hat{X} \gets \hat{X} + \alpha \cdot sign\_gradients$
        \State $\hat{X} \gets \text{clip}(\hat{X}, X - \epsilon, X + \epsilon)$
        \State $\hat{X} \gets \text{clip}(\hat{X}, 0, 1)$
    \EndFor
    \State \Return $\hat{X}$ of size $M$
\EndFunction

\end{algorithmic}
\end{algorithm}
\vspace{-5pt}
\end{figure}

\section{AdvQuNN Methodology}
\label{sec:methodology}

In our methodology, we focus on the adaptability of QuNNs by customizing the quanvolutional filter $U(\theta)$, which is crucial for adapting the parametrized quantum circuit, or Ansatz. This versatility led to the development of AdvQuNN, a methodology aimed at evaluating the robustness of QuNNs, paving the way for enhanced methods to robustify QuNNs against various adversarial attacks by fine-tuning components such as gate selection and entanglement settings.

\begin{figure*}
    \centering
    \includegraphics[width=0.84\linewidth]{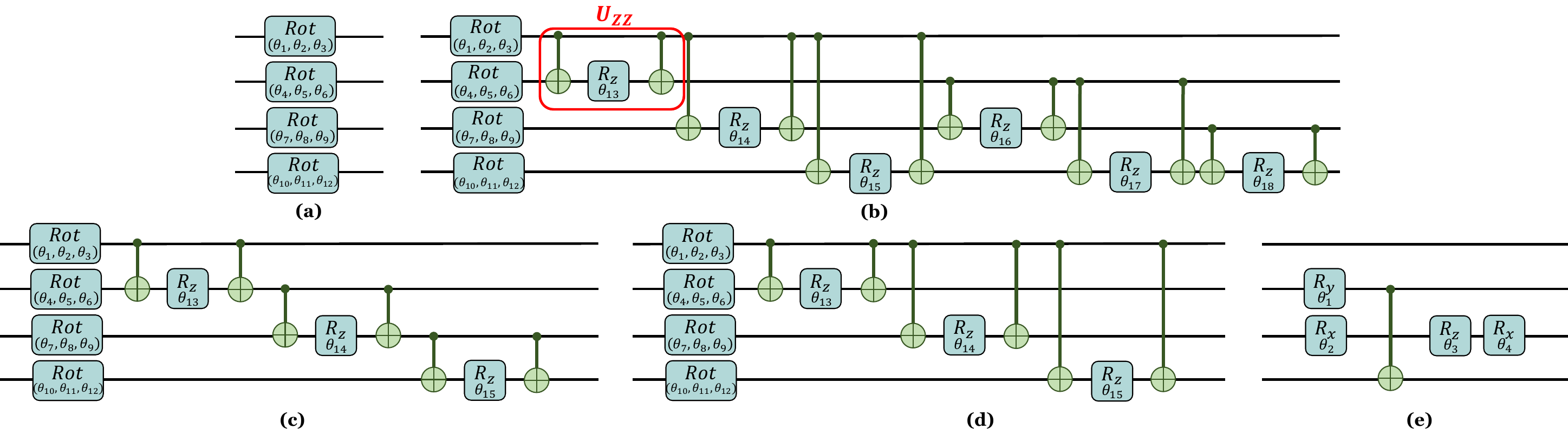}
    \caption{A set of Ansatz architectures considered in the study. Each is labeled as (a) No entanglement, (b) ZZ full entanglement, (c) ZZ linear entanglement, (d) ZZ star entanglement, and (e) Random architecture. The $Rot(a,b,c)$ is a set of three rotation gates $R_z(a)R_y(b)R_z(c)$. The gates $R_x$, $R_y$, and $R_z$ are parameterized.}
    \label{fig:Ansatzes_architectures}
\end{figure*}

Unlike prior research that used random Ansatz architectures, we evaluated various Ansatz architectures and found that the expressibility and entanglement strength of the quantum circuit, influenced by rotation gates and entanglement configurations~\cite{sim2019expressibility}, impact the feature extraction capability of QuNNs, hence its robustness. As quantum entanglement has an important role in the decision boundary formation of the quanvolved data, we explored different entanglement patterns and identified the five most robust architectures for our dataset (Figure~\ref{fig:Ansatzes_architectures}), including no entanglement and random architectures~\cite{henderson2020quanvolutional}, as well as architectures that follow the popular nearest-neighbor and all-to-all entanglement topologies. Additionally, the  circuit gate selection is inspired by data encoding technique proved effective for data separation in Quantum Support Vector Machine (QSVM)~\cite{rebentrost2014quantum}, specifically the $ZZ$ feature map that employs $ZZ$ entanglement.

The $ZZ$ entanglement pattern, involving a controlled phase shift $e^{-i \theta}$ between two qubits. This entanglement is created using the unitary transformation $U_{ZZ}(\theta) = e^{-i \theta Z \otimes Z}$, where $Z$ is the Pauli-$Z$ matrix and $\theta$ is the angle of rotation, as shown in red box Figure~\ref{fig:Ansatzes_architectures}b.

The unitary operators of the tested architectures are as follows:

\textbf{\textit{A) No entanglement architecture (Figure~\ref{fig:Ansatzes_architectures}a).}} it applies a general single-qubit rotation gates to all qubits
    \begin{equation}
        U_{noent}=\bigotimes_0^n Rot(a,b,c)
    \end{equation}
    where $Rot(a,b,c)=R_z(a)R_y(b)R_z(c)$

\textbf{\textit{B) Random architecture (Figure~\ref{fig:Ansatzes_architectures}e).}}: It applies a random circuit.

\textbf{\textit{C) $ZZ$ linear entanglement architecture (Figure~\ref{fig:Ansatzes_architectures}c).}} In this architecture, each qubit is entangled linearly with its nearest neighbor, forming a chain of entangled qubits with just $n+1$ operations for $n$ qubits. Its corresponding unitary operator is:

\begin{equation}
U_{zzlinear}=U^{\mathrm{linear-ent}}\left(\bigotimes_{0}^n Rot\left(a,b,c\right)\right)
\vspace{-10pt}
\end{equation}

\begin{equation}
 U^{\text{linear-ent}}:=\prod_{q=1}^{n-1} \mathbf{E}_{q, q+1} \;\;\;\text{where}\;\;\; \mathbf{E}_{q, q+1}=e^{-i \theta Z_q Z_{q+1}}
\end{equation}

\textbf{\textit{D) $ZZ$ Full Entanglement (Figure~\ref{fig:Ansatzes_architectures}b).}} This architecture involves all-to-all entanglement via $ZZ$ interactions, with entangling gates increasing quadratically, needing $n(n-1)/2$ operations for $n$ qubits. Its corresponding unitary operator is:

\begin{equation}
U_{zzfull}=U^{\mathrm{full-ent}}\left(\bigotimes_{0}^n Rot\left(a,b,c\right)\right)
\end{equation}
\begin{equation}
U^{\text{full-ent }}:=\prod_{q=1}^{n-1} \prod_{k=q+1}^n \mathbf{E}_{q, k} \;\;\;\text{where}\;\;\; \mathbf{E}_{q, k}=e^{-i \theta Z_q Z_k}
\end{equation}

\textbf{\textit{E) ZZ Star Entanglement (Figure~\ref{fig:Ansatzes_architectures}d).}} The Star entanglement pattern links one central qubit to all others, forming a star shape with $n-1$ entangling operations for $n$ qubits, in our case, we center the entanglement on the first qubit. Its corresponding unitary operator is:

\begin{equation}
U_{zzstar}=U^{\mathrm{star-ent}}\left(\bigotimes_{0}^n Rot\left(a,b,c\right)\right)
\vspace{-10pt}
\end{equation}
\begin{equation}
U^{\text{star-ent}}:=\prod_{q=1}^{n-1} \mathbf{E}_{1, q+1} \;\;\;\text{where}\;\;\; \mathbf{E}_{1, q+1}=e^{-i \theta Z_1 Z_{q+1}}
\vspace{-5pt}
\end{equation}



\section{Results and Discussion}

\subsection{Experimental Setup} \label{Experimental_Setup}

\begin{figure}
    \centering
    \includegraphics[width=0.8\linewidth]{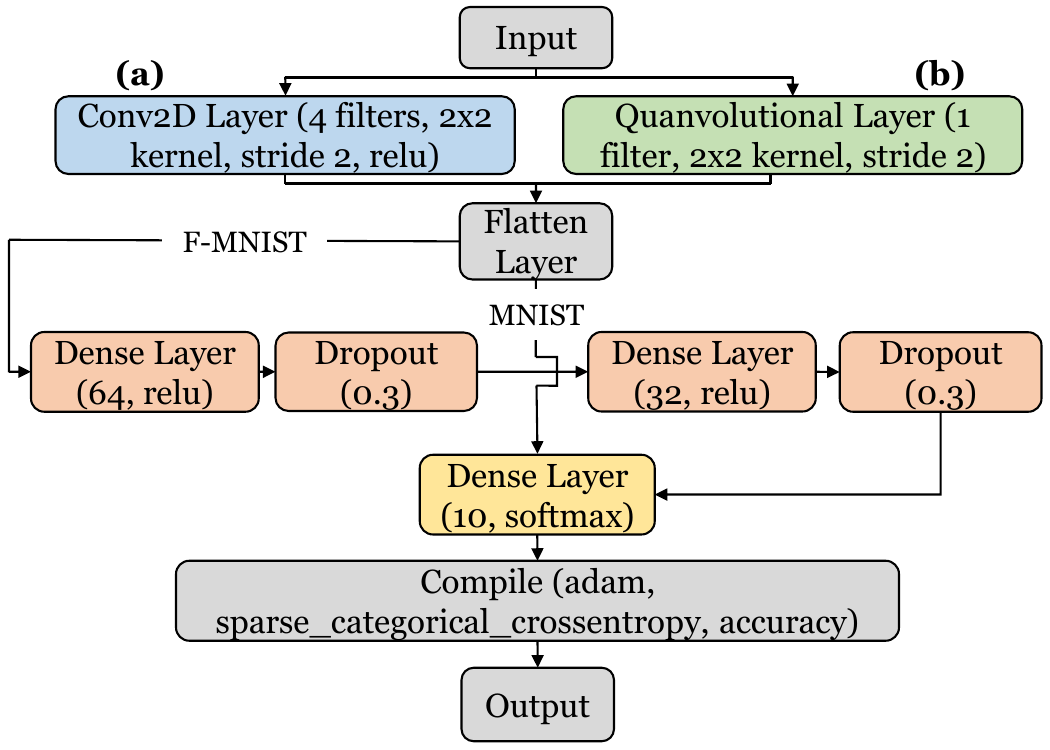}
    \caption{Diagram showing (a) a CNN with a convolutional layer, and (b) a QuNN with a quantum convolutional layer for FMNIST and MNIST datasets. Post-flattening, MNIST data follows a direct path (yellow box) to the softmax layer, while FMNIST data takes a red-box path with dense layers and dropout before softmax. Both paths converge for final model optimization and performance assessment.}
    \label{fig:Flows}
\end{figure}

In our experiment, we evaluated three machine learning models' robustness against Fast Gradient Sign Method (FGSM)~\cite{goodfellow2014explaining}, Projected Gradient Descent (PGD)~\cite{madry2018towards}, and Momentum-based Iterative Method (MIM)~\cite{dong2018boosting} attacks on MNIST~\cite{mnist} and Fashion MNIST (FMNIST)~\cite{FashionMNIST} datasets: two classical models and one quantum model. The first classical model, depicted in Figure~\ref{fig:Flows}a, employs a convolutional layer with a $2\times2$ kernel, stride of $2$, $4$ output channels, and ReLU activation, transforming $28\times28$ pixel images into $14\times14$ feature maps. These maps are then processed by a 10-neuron dense layer with softmax activation. In contrast, the second classical model relies solely on fully connected layers. The third model is the QuNN model with a similar structure to the first classical model for a fair comparison, Figure~\ref{fig:Flows}b.  It includes a non-trainable quanvolutional layer with the same kernel size, stride, and output channels, yielding feature maps of analogous size from qubit expectation values, see Figure~\ref{fig:QNN_architecture}. The fully connected layers configuration in the DNN for MNIST follows a direct path to the yellow box as depicted in Figure~\ref{fig:Flows}. For FMNIST, additional layers are incorporated into the DNN for both classical and quantum models, as shown in Figure~\ref{fig:Flows} red boxes.

For each adversarial attack, we assess the robustness of the QuNNs for each Ansatz architecture in the quanvolutional filter (see Figure~\ref{fig:QNN_architecture}). For each architecture, we run $7$ trials and plot the error bars chart. Each trial involves the following steps:
\begin{enumerate}[leftmargin=*]
    
    

    \item \textbf{Train the QuNN model}: We train and validate the QuNN model using $50$ training samples. Each image is processed through a quanvolutional layer, followed by a classical layer for classification. The model is trained with a batch size of $4$ for $30$ epochs, using a learning rate of $0.001$. The classical layer parameters are updated iteratively.
    
    \item \textbf{Create adversarial examples:} Using the trained model we apply white box FGSM, PGD, and MIM attack on $30$ test images using Algorithms 1, 2, and 3 to create adversarial test images. The perturbations in the image pixels are mapped to rotational perturbations on qubits. This is repeated for varied perturbation strength (epsilon) values. \textit{The values of epsilons are considering minor perturbations to relatively large perturbations to capture the behavior of the models on both spectrums}.

    \item \textbf{Evaluation:} Finally we evaluate the performance of the trained QuNN on adversarial test images at various epsilon values, resulting in accuracy versus epsilon plots.
\end{enumerate}


\subsection{Comparative Analysis between CNNs and QuNNs against Adversarial Attacks}
\label{subsec:comparison_Classical_Quantum}

We conduct experiments on three distinct models as discussed in \Cref{Experimental_Setup}, with a primary focus on the MNIST dataset. Here, we only focus on the ZZ full Ansatz (Figure~\ref{fig:Ansatzes_architectures}b) to demonstrate key insights. Additionally, we conduct further analyses on the FMNIST dataset, highlighting its differences from the MNIST. 

\begin{figure}
    \centering
    \includegraphics[width=0.8\linewidth]{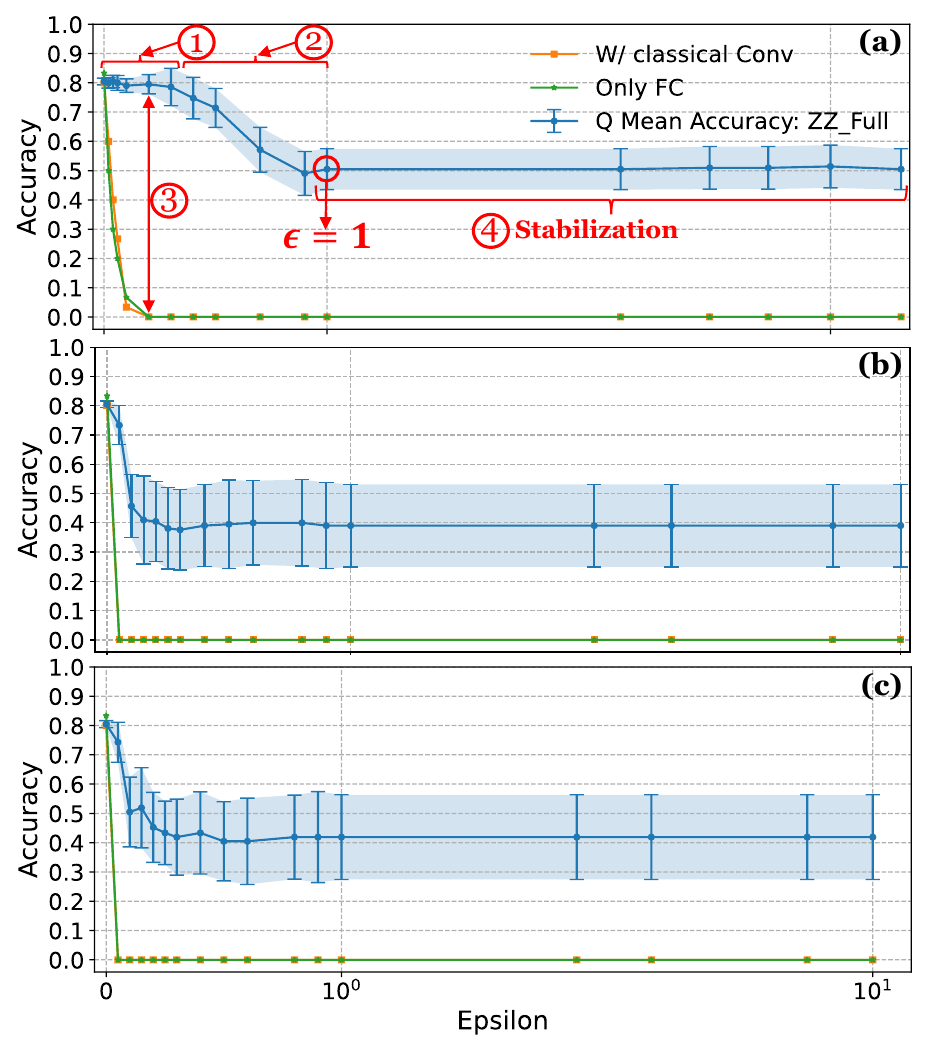}
    \caption{Robustness evaluation of classical and quantum models with ZZ full Ansatz against a) FGSM, b) PGD, and c) MIM adversarial attacks at different perturbation strengths for the MNIST dataset.}
    \label{fig:Q_C_comparison}
\end{figure}

Figure~\ref{fig:Q_C_comparison} presents a comparison of the robustness of the three models when subjected to the FGSM, PGD, and MIM attacks across a range of perturbation strengths, denoted by $\epsilon$. The x-axis represents the magnitude of the adversarial perturbation $\epsilon$, and is set to a logarithmic scale to improve the clarity of the line's pattern at lower levels of perturbation. For the FGSM attack, $\epsilon$ spans from 0 to 15, while for PGD and MIM attacks, it ranges from 0 to 10. The y-axis indicates model accuracy, with values from 0 to 1, where 1 indicates perfect accuracy.

\textbf{A) Results for the FGSM Attack}

In Figure~\ref{fig:Q_C_comparison}a, subjected to FGSM attack, the classical models show a sharp drop in accuracy as $\epsilon$ increases, reaching near zero at smaller perturbations. In contrast, the quantum model with the ZZ full Ansatz retains higher accuracy at low perturbations (pointer \textcircled{1}) and declines more gradually with increasing $\epsilon$ (pointer \textcircled{2}), maintaining over $0.6$ accuracy even at higher values of $\epsilon$ (pointer \textcircled{3}), indicating greater robustness compared to the classical models. The error bars in the quantum model indicate result variability, possibly due to quantum computing's inherent randomness or other factors as discussed in \Cref{remarks}.



\textbf{B) Results for the PGD Attack}

The observations made for the FGSM attack apply similarly to the PGD attack, see Figure~\ref{fig:Q_C_comparison}b. Unlike for the FGSM attack, the classical models under the PGD attack show a quicker decline to low accuracy at lower perturbations due to the enhanced adversarial capability of the PGD attack. The QuNN, despite a decrease in accuracy at small perturbations, maintains higher and more stable accuracy than classical models. The error bars associated with the quantum model are somewhat larger under the PGD attack, implying higher variability in performance, possibly due to the iterative nature of the PGD attack as opposed to the one-step FGSM attack, but it still shows consistent robustness across perturbations. 

\textbf{C) Results for the MIM Attack} 

The performance of the MIM attack is similar to that of the PGD attack, with a slight decrease in accuracy in the QuNN, see Figure~\ref{fig:Q_C_comparison}c. This slight decrease is due to the MIM attack being a more sophisticated attack than the PGD attack. 

For the FMNIST dataset, classical models' accuracy quickly drops to zero at low epsilon values for all attacks. The quantum model also shows reduced accuracy, likely due to FMNIST's greater complexity compared to MNIST. Despite a reduction in accuracy, the model still exhibits an overall trend of higher performance, maintaining an accuracy level around $0.4$, see Figures~\ref{fig:Compare_all_ansatzes}a,~\ref{fig:Compare_all_ansatzes}b, and~\ref{fig:Compare_all_ansatzes}c.

\subsection{Comparative Analysis between Ansatz Architectures} \label{Comparative_Analysis_between Ansatz_Architectures}


\begin{figure*}[t]
    \centering
    \includegraphics[width=\linewidth]{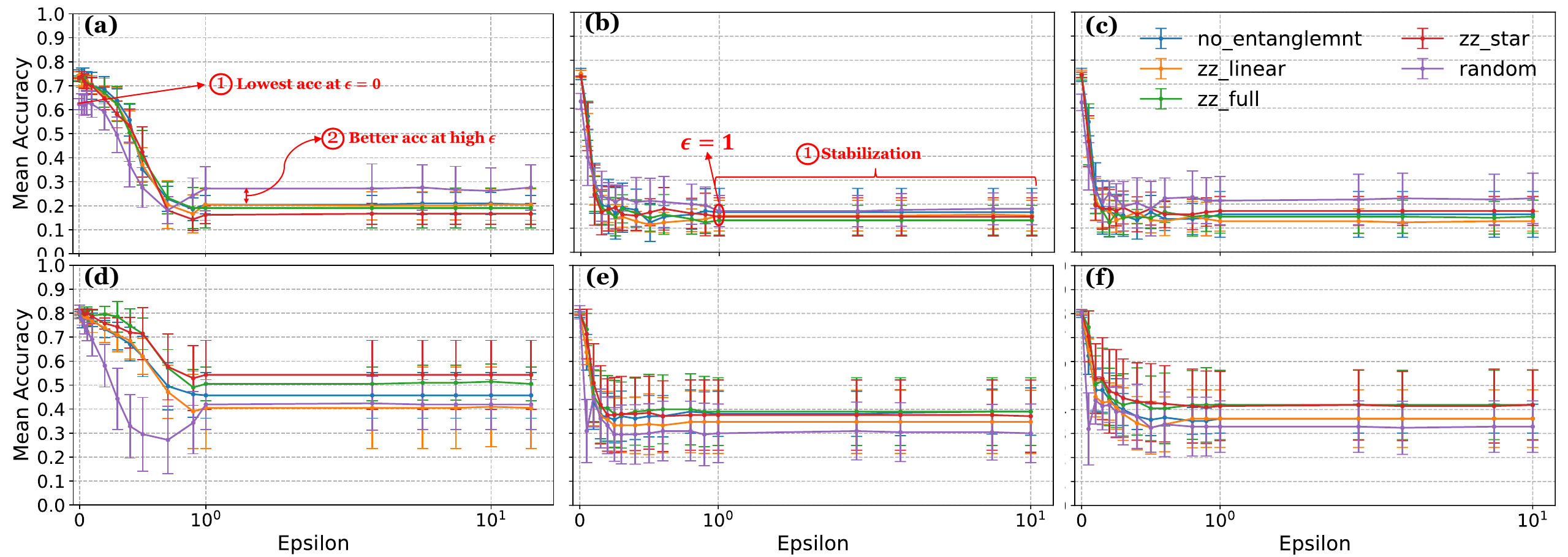}
    \caption{Evaluation of the robustness of the 5 Ansatz architectures under three adversarial attacks: a) FGSM, b) PGD, and c) MIM for the FMNIST dataset and d) FGSM, e) PGD, and f) MIM for the MNIST dataset.}
    \label{fig:Compare_all_ansatzes}
\end{figure*}

In the analysis of adversarial attacks on various quantum circuit architectures, Figures~\ref{fig:Compare_all_ansatzes}d,~\ref{fig:Compare_all_ansatzes}e, and~\ref{fig:Compare_all_ansatzes}f, illustrate the average accuracy of each architecture with error bars for each attack, showing different robustness against adversarial attacks across the MNIST dataset as epsilon varies. The ZZ full and star entanglement architectures exhibit superior robustness compared to all architectures across all three types of attacks (FGSM, PGD, and MIM), maintaining high accuracy even at high perturbation levels. In contrast, the no-entanglement and ZZ linear architectures exhibit better resilience than the Random architecture, showing a more gradual decline in accuracy. On the other hand, the no-entanglement slightly surpasses the ZZ linear and behaves similarly for the MIM attack at high epsilons. Conversely, the Random architecture consistently underperforms, with a sharp accuracy decline at lower perturbations, stabilizing only at higher levels.

With the FMNIST dataset (Figures~\ref{fig:Compare_all_ansatzes}a,~\ref{fig:Compare_all_ansatzes}b, and~\ref{fig:Compare_all_ansatzes}c), the Random architecture outperforms others (pointer \textcircled{2}, Figure\ref{fig:Compare_all_ansatzes}a), despite not achieving the highest accuracy at zero epsilon (pointer \textcircled{1}, Figure~\ref{fig:Compare_all_ansatzes}a). Other architectures show consistent performance across varying epsilon levels and stabilize at high $\epsilon$ (e.g., pointer \textcircled{1}, Figure~\ref{fig:Compare_all_ansatzes}b). Interestingly, the robustness ranking of the Ansatz architecture varies between the FMNIST and MNIST datasets. This difference highlights the impact of the chosen dataset on the Ansatz architecture's performance. It suggests a strong link between the robustness of the QuNN model and the configuration of its quantum circuit, which is dependent on the specific dataset being employed.


\subsection{Remarks and discussion} \label{remarks}

The QuNN circuit includes angle encoding combined with a circuit that exhibits various entanglement scheme patterns. While angle encoding alone can increase robustness due to better decision boundary separation~\cite{larose2020robust, west2023towards}, at low epsilons, this combination enhances the QuNN's resilience against adversarial attacks as the circuit's entanglement increases. This is evident in the results for the MNIST dataset, where the transition from no entanglement (Figure~\ref{fig:Ansatzes_architectures}a) to full entanglement (Figure~\ref{fig:Ansatzes_architectures}b) demonstrates improved resilience. Recent theoretical studies~\cite{dowling2024adversarial} have shown that angle encoding, which involves local operations, when combined with a quantum circuit exhibiting quantum chaos (high local-operator entanglement), can provide robustness against universal adversarial attacks. Our work provides empirical proof for this theoretical study.

All Ansatz architectures show a plateau in accuracy when perturbation strength exceeds $1$ (see pointer \textcircled{4}, Figure~\ref{fig:Q_C_comparison}a). This plateau occurs because image pixel values are clipped between 0 and 1, which makes all attacked images look identical beyond epsilon equal to 1. Consequently, the QuNN performance is the same. This accuracy varies for each model training run and adversarial example generation, as indicated by the error bar lengths. Since FGSM, PGD, and MIM attacks use gradient information that varies across training sessions, the adversarialness of examples differs, affecting the interaction of the fixed quanvolutional layer with these images and resulting in variable error bar sizes.

The non-trainable quanvolutional layer in QuNN acts as a low-pass filter, reducing extreme perturbations and retaining robust, meaningful features of the image, whereas classical models depend on informative but vulnerable features~\cite{west2023benchmarking}, which explains the significant difference in performance. As a result, QuNN's performance only decreases significantly under very high perturbations. 
If this quanvolutional filter is designed adequately, reflecting the architecture of the Ansatz, it may enhance the extraction of robust features from the input images. For instance, for the MNIST dataset, a common characteristic between the $ZZ$ full and $ZZ$ star architectures, both entanglement patterns involve the first qubit getting entangled with all others. We can seek such characteristics to design tailored Ansatz architecture specific to the dataset.

\section{Conclusion}

The AdvQuNN analysis reveals key properties that enhance QuNN model robustness against adversarial attacks. By examining various Ansatz architectures, it is evident that qubit connectivity patterns significantly influence data extraction and robustness. Specifically, for the MNIST dataset, increased qubit entanglement allows the model to detect and learn robust features, making it harder for adversaries to introduce effective perturbations. This confirms the experiment results where QuNNs require more significant perturbations to experience a decrease in accuracy, whereas the classical models are susceptible to accuracy loss even with minor perturbations. Despite confirming recent claims about the robustness of QuNNs, this study goes a step further by demonstrating their relativity to the quantum circuit design, marking this study the first research to uncover these important insights.

In future research, we aim to extend our findings to various QML models across more datasets and attack scenarios. We plan to design specialized, data- and attack-oriented quantum circuits to enhance robustness, delve into QuNN accuracy trends, and explore strategies to minimize error margins. This study paves new pathways for robustifying QuNN systems by custom-designing quantum circuit architectures to counter adversarial attacks, embedding inherent robustness directly into the model's structure, representing a significant shift in improving QML resilience against threats.

\section*{Acknowledgment}
This work was supported in parts by the NYUAD Center for Quantum and Topological Systems (CQTS), funded by Tamkeen under the NYUAD Research Institute grant CG008, and the NYUAD Center for Cyber Security (CCS), funded by Tamkeen under the NYUAD Research Institute Award G1104.

\newpage

\bibliographystyle{ieeetr}
\bibliography{main.bib}

\end{document}